\documentclass[twocolumn,prl,aps,groupedaddress,nopacs,footinbib,preprintnumbers,floats,floatfix]{revtex4}

\usepackage{amsmath}
\usepackage{amssymb}
\usepackage{latexsym}
\usepackage{graphicx}
\usepackage{hyperref}
\usepackage{mathrsfs}
\usepackage{bm}

\newcommand{\be}{\begin{equation}}
\newcommand{\ee}{\end{equation}}
\def\bea{\begin{eqnarray}}
\def\eea{\end{eqnarray}}
\def\ba{\begin{eqnarray}}
\def\ea{\end{eqnarray}}

\newcommand{\ls}{{\mathscr{L}}}

\newcommand{\bras}[1]{\left[#1\right]}

\begin{document}

\title{A litmus test for $\Lambda$}

\author{ Caroline Zunckel$^1$ \& Chris Clarkson$^2$  \\
\it $^1$ Oxford University, Astrophysics,  Denys Wilkinson Building, Keble
Road, OX1 3RH, United Kingdom\\
\it $^2$Cosmology \& Gravity Group, Department of Mathematics
and Applied Mathematics, University of Cape Town, Rondebosch 7701, Cape Town,
South Africa}

\begin{abstract}
The critical issue in cosmology today lies in determining if the cosmological constant is the underlying ingredient of dark energy. Our profound lack of understanding of the physics of dark energy places severe constrains on our ability to say anything about its possible dynamical nature. Quoted errors on the equation of state, $w(z)$, are so heavily dependent on necessarily over-simplified parameterisations they are at risk of being rendered meaningless. Moreover, the existence of degeneracies between the reconstructed $w(z)$ and the matter and curvature densities weakens any conclusions still further. We propose consistency tests for the cosmological constant which provide a direct observational signal if $\Lambda$ is wrong, regardless of the densities of matter and curvature. As an example of its utility, our flat case test can warn of a small transition from $w(z)=-1$ of $20\%$ from SNAP quality data at 4-$\sigma$, even when direct reconstruction techniques see virtually no evidence for deviation from~$\Lambda$. It is shown to successfully rule out a wide range of non-$\Lambda$ dark energy models with no reliance on knowledge of $\Omega_m$ using SNAP-quaility data and a large range for using $10^5$ supernovae as forecasted for LSST.
\end{abstract}

\maketitle

\paragraph{Introduction}
The discovery that the expansion rate of the universe is apparently speeding up from the type-1a supernovae is arguably one of the most significant events in the history of modern cosmology \cite{review}. Although seemingly consistent with our current concordance model in which the source of the cosmic acceleration, coined `dark energy',  takes on the form of Einstein's cosmological constant, the precision of current data is not sufficient to rule out the possibility of an evolving component. A huge question mark still exists: whether the canonical flat $\Lambda$CDM model is correct or not. If not then we are perhaps looking for some dynamical field with a repulsive gravitation force, or maybe the field equations are wrong on Hubble scales. More bizarrely, this could instead be indicating that the Copernican principle is wrong, and that radial inhomogeneity is responsible for our confusion~\cite{CP}. Within the FLRW paradigm, all possibilities can be characterised, as far as the background dynamics are concerned, by the dark energy equation of state $w(z)$. Unfortunately, from a theoretical perspective we have virtually nothing to go on, implying that $w(z)$ could really be pretty much anything. Our priority in cosmology today must therefore lie in searching for evidence for $w(z)\neq-1$.

The observational challenge to such ambiguity lies in trying to find a straightforward, yet meaningful and sufficiently general way to treat $w(z)$. 
This is usually in terms of a simple parameterisation of $w(z)$; but any such functional forms for $w(z)$ are highly problematic because they have no basis in a sound physical theory and to be flexible require a set of parameters that becomes far too large to be meaningful. Furthermore, they are only sensitive to deviations
from $w=-1$ in a highly limited function space~\cite{BCK}. To try to reduce the huge arbitrariness, the space of allowed $w(z)$ models is often reduced on physical grounds 
to $w\geq-1$; but if $w$ is only effective, parameterising a modified gravity model, say, then this might be too limiting. Perhaps the dark energy is getting up to all-sorts: until we have a way of predicting this, prematurely deciding that $w(z)$ must be slowly changing or undergoing a sharp transition say, leaves vast unchartered blind spots in the possible function space of dark energy~\cite{BCK}.  
Ideally, then, we need a non-parametric way of determining the functional form of $w(z)$.

An alternative procedure is to reconstruct $w(z)$ directly from the observables without any dependence on a parameterisation of $w(z)$ or understanding of dark energy~\cite{weller_albrecht,saini,chiba}.
Such direct reconstruction methods rely on estimating the first and 
second derivatives of luminosity-distance data. Errors on $d_L(z)$ alone do 
not translate simply to errors on $w(z)$: $d_L(z)$ can deviate from $\Lambda$CDM  $<1\%$ even though $w(z)$ is fluctuating wildly. The reverse of this was shown by means of striking plots in \cite{maor}. 
Furthermore, the precision to which we can estimate the matter and curvature densities have acute ramifications for estimates of 
$w(z)$~\cite{Kunz,CCB,maor,shafieloo}. 

Because the quality of the current data is not yet of sufficient quality to determine the exact form of $w(z)$ or to distinguish between different models with much confidence, it is difficult to be satisfied that $w=-1$. Here, we demonstrate a method to put the concordance model to the test by introducing a simple null test for $\Lambda$CDM directly from distance-redshift data. We construct a relation $\ls(z)$ which has the property of being exactly equal to 0 over any redshift range of the reconstruction only for $w=-1$. Critically, it is independent of the matter density $\Omega_m$. Using the Pad\`{e} ansatz introdued in \cite{saini} to fit directly to the luminosity-distance curve, we show that $\ls(z)$ can be used to 
rule out $w=-1$ for a wide range of types of evolving $w(z)$ for $10^5$ SN as predicted by LSST \cite{LSST}, with certain test cases of $w(z)$ 
being ruled out using only SNAP-like data. We then extend our test to include curvature, thereby providing a genuine litmus test for the cosmological constant. 

\begin{widetext}
\paragraph{Reconstructing Dark Energy}

The dark energy equation of state, $w(z)$, is typically reconstructed using distance measurements as a function of redshift. 
The luminosity distance may be written as 
\begin{equation}\label{d_L}
d_{L}(z)=\frac{c(1+z)}{H_0 \sqrt{-\Omega_k}}\sin{\left( 
\sqrt{-\Omega_k}\int_0^z{\mathrm{d}z'\frac{H_0}{H(z')}}\right)},
\end{equation}
which is formally valid for all curvatures, where $H(z)$ is given by the Friedmann equation, 
\be
H(z)^2= H_0^2\biggl\{\Omega_{m} (1+z)^3+\Omega_{
k}(1+z)^2
+\Omega_{DE}\exp{\left[3\int_0^z
\frac{1+w(z')}{1+z'}\mathrm{d}z'\right]}\biggr\},
\ee
and $\Omega_{DE}=1-\Omega_m-\Omega_k$. The usual procedure is to postulate a several parameter form for $w(z)$ and calculate $d_L(z)$ using Eq.~(\ref{d_L}).  
An alternative method is to reconstruct $w(z)$ by directly reconstructing the luminosity-distance curve.  It has been shown in~\cite{astier, CCB} $d_L(z)$ 
may be inverted to yield $w(z)$. Writing $D(z)=(H_0/c)(1+z)^{-1}d_L(z)$, we have 
\be
w(z)=\frac{2(1+z)(1+\Omega_kD^2)D''-\left[(1+z)^2\Omega_kD'^2+2(1+z)\Omega_kDD'-3(1+\Omega_kD^2)\right]D'}
{3\left\{(1+z)^2\left[\Omega_k+(1+z)\Omega_m\right]D'^2-(1+\Omega_kD^2)\right\}D'}.
\label{w}
\ee
Thus, given a distance-redshift curve $D(z)$, we can reconstruct the dark energy equation of state~\cite{reviewofreconstruction}. Typically, one chooses a parameterised ansatz for $D(z)$, such as the Pade ansatz used in~\cite{saini},
\be
D_L(z) = \bras{\frac{(1+z) - \alpha \sqrt{(1+z)}-1+\alpha}{\beta(1+z)+\gamma \sqrt{1+z}+2-\alpha-\beta-\gamma}}.
\label{pade}
\ee
one then fits it to the data, and then calculates $w(z)$ from Eq~(\ref{w}). Such a reconstruction method is more generic than parameterising $w(z)$ directly because we are fitting directly 
to the observable, and so can spot small variations in $d_L$ which can translate to dramatic variations in $w(z)$. Unfortunately, this also leads to larger reported errors in $w(z)$ than calculated by parameterising it directly. It is not clear, however, which errors on  our understanding of dark energy should be taken most seriously. What is very nice about this method is that, if done in small redshift bins, measurements of $w(z)$ in a given bin is independent of bins at lower $z$; this is not the case for parameterised $w$ methods as they integrate over redshift when calculating $d_L(z)$. Both methods suffer from strong degeneracies with $\Omega_k$ and $\Omega_m$, however~-- see, for example \cite{CCB,Kunz,shafieloo}. This vital problem we circumvent using the tests we present below.
\vspace{5mm}
\end{widetext}
 \paragraph{Consistency Test for $\Lambda$CDM} For flat $\Lambda$CDM models the slope of the distance data curve must satisfy $D'(z)=1/\sqrt{\Omega_m(1+z)^3+(1-\Omega_m)}$. Rearranging for $\Omega_m$ we have
\be\label{Om}
\Omega_m=\frac{1-D'(z)^2}{[(1+z)^3-1]D'(z)^2}.
\ee
Within the flat $\Lambda$CDM paradigm, if we measure $D'(z)$ at some $z$ and calculate the rhs of this equation, we should obtain the same answer independently of the redshift of measurement. Differentiating Eq.~(\ref{Om}) we then find that
\ba
\label{eq:L}
\mathscr{L}(z)\!\!&=&\!\!\zeta D''(z)+3(1+z)^2D'(z)[1-D'(z)^2]\nonumber\\
&=&\!\!0~\mathrm{for~all~flat}~\Lambda\mathrm{CDM~models}.
\ea
We have used the shorthand $\zeta=2[(1+z)^3-1]$.
\emph{Note that this is completely independent of the value of $\Omega_m$.} We may use this as a test for $\Lambda$ as follows: take a parameterised form for $D(z)$ and fit to the data. If $\ls(z)=0$ lies outside the n-$\sigma$ error bars, we have n-$\sigma$ evidence for deviations from $\Lambda$, assuming the curvature is zero (see below for the generalised test without this restriction). If, on the other hand, $\ls(z)=0$  is a sufficiently good fit for all suitable parameterisation we can think of, then that is good evidence for $\Lambda$. But the important thing is that only \emph{one choice of parameterisation has to imply $\ls(z)\neq0$ to provide evidence that $\Lambda$CDM is wrong}. Every parameterisation has many blind spots~-- variations in $w(z)$ that could never be picked up~-- so by searching through various ensures that the blind spots are illuminated. 
\begin{figure*}[t!]
\begin{center}
\includegraphics[width=1.0\textwidth]{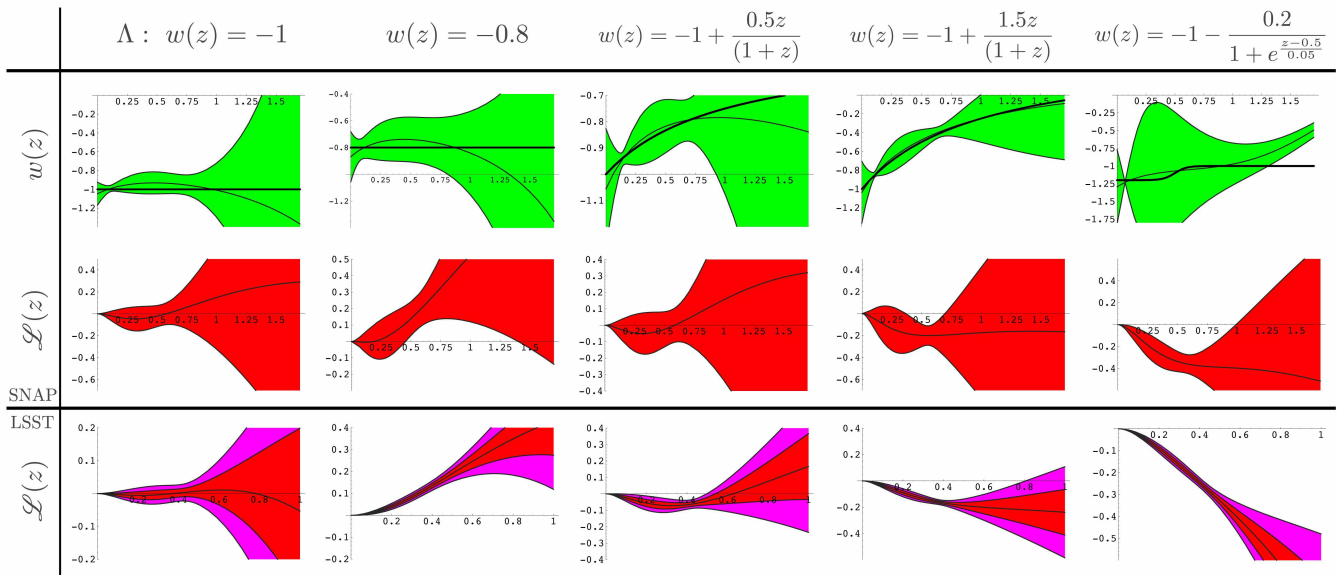} 
\caption{\emph{Consistency test for $\Lambda$}. We show $\ls(z)$ for various dark energy models that we could expect to reconstruct from SNAP (middle row) and LSST )bottom row) data, which is independent of $\Omega_m$. Also shown is the reconstructed $w(z)$ using the Pad\`{e}  ansatz, Eq.~(\ref{pade}), together with 1-$\sigma$ errors; this reconstruction assumes the correct choice of $\Omega_m$, and it considerably worse without this. For SNAP data, we see that $\ls(z)$ always provides evidence for evolving dark energy, except in the case when it is evolving slowly, as in the middle column. Note that the Pad\`{e}  ansatz is especially well adapted for an evolving model such as that in the fourth column, but is extremely poor for constant $w$, or where it undergoes a sharp transition (fifth column). Indeed, in the transition model, it provides little evidence for $w\neq-1$ while $\ls(z)$ signals something is happening at 4-$\sigma$. In the third row, we see that $\ls(z)$ will be able to signal evolving dark energy to high confidence using LSST data. (Note the different redshift ranges used for SNAP and LSST.)}
\label{fig}
\end{center}
\end{figure*}

\paragraph{Testing the test} To illustrate the test, and to demonstrate that it will work, we fit the Pad\`{e} ansatz given in Eq.~(\ref{pade}) to the luminosity-distance data for a set of simulated data. The derivatives $D{''}$ and $D{'}$ are then calculated and substituted into Eq.~(\ref{eq:L}). A deviation of $\ls(z)=0$ within the error bars of the reconstruction signals the detection of a non-$\Lambda$ paradigm. In order to test the consistency equation, we select a sub-class of parameterizations encompassing a wide class of of DE models. In addition to $w=-1$, we consider quintessence-type models with $w=$constant, slowly evolving $w$ models, and a model which undergoes a rapid transition.

To begin with we use simulated supernovae data which we expect to have from the SNAP mission~\cite{SNAP}. The results for $\mathscr{L}(z)$ (in red) for $w(z)$ cases discussed above from 1000 realizations of SNAP-quality data with with $2013$ SN distributed throughout $z=0-1.7$ as done in \cite{shafieloo} with Gaussian noise are shown in the top two rows of Fig.~\ref{fig}.

We find that with SNAP data we will be able to rule out $\Lambda$CDM to more than 1$\sigma$ in certain redshift ranges for most dynamical models, whereas $\mathscr{L}(z)=0$ is included in the error bars for $w=-1$ for all $z$ as we would expect. 
The discrepancies in the effectiveness of $\ls(z)$ for different models is readily understood. The example $w(z)$ in the third and fourth columns represent dark energy models in which the equation of state evolves away from $w=-1$ with increasing $z$. We know that dark energy with $w<0$ becomes less dominant at earlier times which is why its impact on the observables registers to a lessening degree with increased $z$. The reconstructed $w(z)$ for each case is shown in green in the top row. Although the reconstruction of $w(z)$ itself appears in some cases to be more effective in ruling out $\Lambda$, the results are heavily reliant of the \emph{correct} choice of $\Omega_m$. Perturbing $\Omega_m$ used in the reconstruction by only $7\%$
from the true value of the simulated data renders an incorrect reconstruction and includes $w=-1$ with the $1\sigma$ error bars. $\mathscr{L}(z)$ however is unaffected by $\Omega_m$.  Furthermore, we can see that in some cases the two methods can be complimentary: for the case $w=-0.8$, we see that the standard reconstruction rules out $\Lambda$ to $>$1-$\sigma$ for $z<0.75$, while $\ls(z)$ rules it out for $0.5<z<1.3$.

\begin{figure}
\begin{center}
\includegraphics[width=0.8\columnwidth]{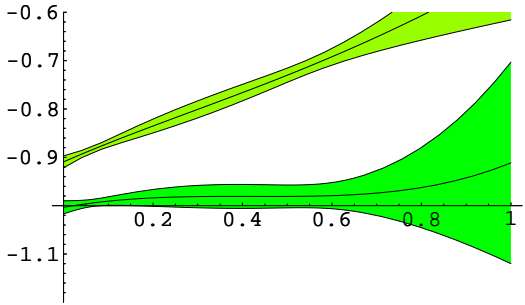} 
\caption{Reconstructed $w(z)$ for $w=-1$ when correct $\Omega_m=0.275$ is assumed (dark green) and when the incorrect value $\Omega_m=0.2$ is assumed (light green) with 1$\sigma$ error bars.
This is for a data set of $10^5$ SN as predicted from LSST. This demonstrates the short-comings of reconstruction methods which rely on precise measurements of the matter density. Our consistency test is immune to such pitfalls.  }
\label{wrong}
\end{center}
\end{figure}

The information that can be derived from a SNAP-like data set is not of a sufficient accuracy to rule out $w=-1$ for all of our test cases.  
We now consider the requirements of a future-generation survey to determine if this is a robust test of $\Lambda$. We consider a future data set of $10^5$ SN of SNAP-like quality  distributed uniformly throughout $z=0-1$ in line with predictions from LSST \cite{LSST}. Figure \ref{fig} shows $\ls(z)$ for the same set of models 
onsidered above, in the bottom row. It is clear that $w=-1$ is ruled out to at least $2\sigma$ in all cases. Although $10^5$ SN may be regarded as a somewhat optimistic estimate, it shows the potential of our consistency relation to test for $\Lambda$ without knowing $\Omega_m$ at all. Even though the reconstruction of $w(z)$ will improve with such a large data set, its innate dependence on the matter density $\Omega_m$ and the method of reconstruction means that the error bars although smaller are not the whole story. This point is illustrated in Fig.~\ref{wrong}, where we contrast the reconstructed $w(z)$ for $10^5$ SN when we assume the correct $\Omega_m=0.275$ and the wrong $\Omega_m = 0.2$.

\paragraph{Including curvature}
Our consistency test may easily be extended to include curvature as follows: with $\Omega_k\neq0$, $\ls(z)$ becomes:
\ba
\ls(z)\!\!&=&\!\!\left\{\zeta D^2D''-[z^2(3+z)(1+z)D'^2+\zeta DD'\right.\nonumber\\
\!\!&-&\!\!\left.3(1+z)^2D^2]D'\right\}\Omega_k
+\zeta D''+3(1+z)^2D'(1-D'^2) \nonumber\\
\!\!&\equiv&\!\!0~~~\mbox{for} ~~~\Lambda\mathrm{CDM}.\label{L_gen}
\ea
In order to formulate a test which is independent of curvature, we can rearrange this for $\Omega_k$ and differentiate it to yield a (large!) expression whose numerator must be equal to zero. Unfortunately, as it involves third derivatives of $D(z)$, it doesn't appear to be very useful: an analysis similar to the above reveals that even $10^5$ SN is not enough. Rearranging for $\Omega_k$ and comparing at different redshifts will also be a test for $\Lambda$. On the other hand, curvature may be determined directly from $D(z)$ and $H(z)$ data using $\Omega_k=\{\left[H(z)D'(z)\right]^2-1\}/{[H_0D(z)]^2}$~\cite{CCB}. Substituting this into Eq.~(\ref{L_gen}) gives an observational litmus test for $\Lambda$ which is independent of all cosmological density parameters: 
\ba
\ls_\mathrm{gen}(z)\!\!&=&\!\!\left\{\zeta D^2D''-[z^2(3+z)(1+z)D'^2\right.\nonumber\\
\!\!&+&\!\!\left.\zeta DD'-3(1+z)^2D^2]D'\right\}h(z)^2+\zeta  D\nonumber\\
\!\!&-&\!\!3(1+z)[3(1+z)D^2-z^2(z+3)]D',
\ea
where $h(z)=H(z)/H_0$. Given $H(z)$ data from BAO measurements or from relative age measurements of passively evolving galaxies, we can expect this more general test to be just as useful as the simplified version we have analysed here by the time SNAP or LSST data is available. Note that it involves only second derivatives of distance data and no derivatives of the Hubble rate.  
\paragraph{Conclusions}
We have proposed a simple and direct litmus test for the canonical $\Lambda$CDM paradigm. Our flat consistency test is shown to rule out $w=-1$ for a broad range of dynamical models of dark energy to at least $2\sigma$ using $10^5$ SN as forecasted for LSST while being entirely independent of the matter density. We have also shown that rapid variations in $w(z)$ can be constrained extremely tightly by data we can expect very soon. We have also proposed a litmus test for the cosmological constant which is independent of all cosmological parameters. This may play an important role in determining the true nature of the dark energy. 
\paragraph{Acknowledgements}
We would like to thank
Pedro Ferreira, Roy Maartens, Joe Zuntz and Mark Sullivan for useful comments. CC acknowledges support from the NRF (South Africa). CZ is supported by a Domus A scholarship awarded by Merton College.


\begin{references}

 \bibitem{review}
See, e.g., Friedman, J.A., Turner, M. and Huterer, D.  arxiv:0803.0982 (2008)

\bibitem{CP}
Clarkson, C., Bassett, B. and Lu, T. H.-C. 
	Phys. Rev. Lett., 101 011301 (2008).  See also C{\'e}l{\'e}rier, 
M.-N. arXiv:astro-ph/0702416 (2007) for a review.

\bibitem{BCK} Bassett, B.~A., 
Corasaniti, P.~S., \& Kunz, M.\ 2004, ApJ Lett., 617, L1 

\bibitem{weller_albrecht}
Weller J.,  Albrecht A.,  2002, Phys. Rev., D65, 103512

\bibitem{saini}
Saini, T.D \emph{et al.}, 2000, Phys. Rev. Lett., 85, 1162.

\bibitem{chiba}
Chiba, T. \& Nakumura, T., 2000, Phys. Rev.D. 62, 121301

\bibitem{maor}
Maor, I., Brunstein, R. \& Steinhardt, P., 2001, Phys. Rev. Lett. 86, 6

\bibitem{Kunz}
Kunz, M. arXiv:astro-ph/0702615 (2007)

\bibitem{CCB}
Clarkson, C., Cort\^ez, M. \& Bassett, B. A. JCAP08(2007)011

\bibitem{shafieloo}
Shafieloo, A. \emph{et al.}, 2006, Mon. Not. Roy. Astron. Soc., 366, 1081

\bibitem{LSST}
Tyson, J.A. \& LSST Collaboration, astro-ph/0609516 (2006)

 \bibitem{SNAP}
Aldering G.,  \emph{et~al.}, 2004, astro-ph/0405232
\end{references}
\end{document}